\documentclass[a4paper,10pt]{article}

\usepackage{amssymb}
\usepackage{amsfonts}
\usepackage{amsmath}
\usepackage{subfigure}
\usepackage{algorithm}
\usepackage[noend]{algorithmic}
\usepackage{amsmath,amssymb,cite}
\usepackage{epsfig}
\usepackage{pst-node,pstricks}

\usepackage{graphicx}
\usepackage{psfrag}
\usepackage[nohead, top=2.9cm, bottom=3.2cm, left=2.9cm, right=3.1cm]{geometry}

\def\q5uad{\quad\quad\quad\quad\quad}

\title{Two RPG Flow-graphs for Software Watermarking using Bitonic Sequences of Self-inverting Permutations  \vspace{0.2cm}}

\author{Anna~Mpanti \ \ \ Stavros~D.~Nikolopoulos}

\date{}

\begin{document}

\maketitle

\vspace{-0.5cm}

\centerline{\it Department of Computer Science \& Engineering}

\centerline{\it University of Ioannina}

\centerline{\it GR-45110 \ Ioannina, Greece}

\centerline{\tt \{ampanti,stavros\}@cs.uoi.gr}


\vskip 0.3in

\begin{abstract}
\noindent Software watermarking has received considerable attention and
was adopted by the software development community as a technique to prevent or discourage software piracy and copyright infringement.
A wide range of software watermarking techniques has been proposed among which the graph-based methods
that encode watermarks as graph structures.
Following up on our recently proposed methods for encoding watermark numbers $w$ as reducible permutation flow-graphs $F[\pi^*]$ through the use of self-inverting permutations $\pi^*$, in this paper, we extend the types of flow-graphs available for software watermarking by proposing two different reducible permutation flow-graphs $F_1[\pi^*]$ and $F_2[\pi^*]$ incorporating important properties which are derived from the bitonic subsequences composing 
the self-inverting permutation $\pi^*$. We show that a self-inverting permutation $\pi^*$ can be efficiently encoded into either $F_1[\pi^*]$ or $F_2[\pi^*]$ and also efficiently decoded from theses graph structures.
The proposed flow-graphs $F_1[\pi^*]$ and $F_2[\pi^*]$ enrich the repository of graphs which can encode the same watermark number $w$ and, thus, enable us to embed multiple copies of the same watermark $w$ into an application program $P$. Moreover, the enrichment of that repository with new flow-graphs increases our ability to select a graph structure more similar to the structure of a given application program $P$ thereby enhancing the resilience of our codec system to attacks.

\vspace*{0.1in}
\end{abstract}

\vspace*{0.2in}
\section{Introduction}
\vspace*{0.05in}

Over the last 25 years, digital or multimedia watermarking has become a popular technique for protecting the intellectual property of any digital content such as image, audio, video or software data. In this domain, software watermarking has received considerable attention and was adopted by the software development community as a technique to prevent or discourage software piracy and copyright infringement. A wide range of software watermarking techniques has been proposed among which the graph-based methods that encode watermark numbers as graphs whose structure resembles that of real program graphs.

\vspace*{0.15in}
\noindent {\bf Watermarking}. Digital watermarking is a popular technique for copyright protection of a digital object or, in general, multimedia information \cite{Sa,TaNaMoMa}; the idea is simple: a unique marker, which is called watermark, is embedded into a digital object which may be used to verify its authenticity or the identity of its owners \cite{CoNa,Gro}. The software watermarking problem can be described as the problem of embeding a structure $w$ into a program $P$ such that $w$ can be reliably located and extracted from $P$ even after $P$ has been subjected to code transformations such as translation, optimization and obfuscation \cite{CN12,MyCo}. More precisely, given a program $P$, a watermark $w$, and a key $k$, the software watermarking problem can be formally described by the following two functions: $embed(P, w, k)\to P$ and $extract(P_w, k)\to w$.

Since the late 1990s, there has been an explosion in the number of digital watermarking techniques among which time-series, biological sequences, graph-structured data, spatial data, spatiotemporal data, data-streams and others \cite{PaScSeBe}. Recently, software watermarking has received considerable attention and many researchers have developed several codec algorithms mostly for watermarks that are encoded as graph-structures \cite{EpGoLaMaMiTo}. The patent by Davidson and Myhrvold \cite{DaMy} presented the first published software watermarking algorithm. The preliminary concepts of software watermarking also appeared in paper \cite{GhHe} and patents \cite{MoCo,Sa}. Collberg et al. \cite{CoTho,CoThoLo} presented detailed definitions for software watermarking. Authors of papers \cite{ZhYaNiNi,ZhThWa} have given brief surveys of software watermarking research.

\vspace*{0.15in}
\noindent {\bf Our Contribution}.  Recently, we have presented codec algorithms, namely ${\tt Encode\_W.to.SIP}$ and ${\tt Decode\_SIP.to.W}$, for encoding an integer $w$ into a self-inverting permutation $\pi^*$ and extracting it from $\pi^*$ \cite{CN10}, and several codec algorithms for encoding $\pi^*$ into many reducible permutation flow-graphs $F_i[\pi^*]$ ($i>1$) \cite{CN11,CN15}, having thus created a wide repository of graph-structures, namely flow-graphs, whose structures resemble that of real program graphs.

In this paper, we extend the types of flow-graphs which can efficiently encode a self-inverting permutation $\pi^*$ by proposing two different reducible permutation flow-graphs $F_1[\pi^*]$ and $F_2[\pi^*]$ having properties which are derived from the bitonic subsequences $b_1^*, b_2^*, \cdots, b_k^*$ composing the self-inverting permutation $\pi^*$. We show relations between the elements of such a bitonic subsequence $b_i^*$ and their indices in $\pi^*$ and prove properties for the first, last, max and min elements of $\pi^*$. We also show that the first bitonic subsequence $b_1^*$ of a self-inverting permutation $\pi^*$ of length $n^*$ has always length $\lceil n^*/2 \rceil$ and structure $(\lceil n^*/2 \rceil, \ldots, \pi^*_{max}, \ldots, 1)$, where $\pi^*_{max}$ is the max element of $\pi^*$.

Taking advantage of these properties, we construct two different reducible permutation flow-graphs $F_1[\pi^*]$ and $F_2[\pi^*]$ which can encode the same self-inverting permutation $\pi^*$ and thus, the same watermark number $w$. By construction, the indegree of the first node $s=u_{n^*+1}$ of the flow-graph $F_1[\pi^*]$ is equal to the number of bitonic subsequences $b_1^*, b_2^*, \cdots, b_k^*$ of $\pi^*$, while the indegree of the first node of the graph $F_2[\pi^*]$ is much smaller that $k$. This property causes $F_2[\pi^*]$ more appropriate, in same cases, since it does not contain an extreme characteristic thereby enhancing the resilience of graph-structure to attacks.

The flow-graphs $F_1[\pi^*]$ and $F_2[\pi^*]$ enrich the repository of graphs which can encode the same watermark number $w$ and, thus, enable us to embed several copies of the same watermark $w$ into an application program $P$. Moreover, it increases our ability to select a graph structure more similar to the structure of a given application program $P$ thereby enhancing the resilience of our codec system to attacks.


\vspace*{0.15in}
\noindent {\bf Road Map}. The paper is organized as follows: In Section~2 we establish the notation and related terminology, we present background results, and show properties of the bitonic subsequences which compose a self-inverting permutation $\pi^*$. In Sections~3 and 4 we present our two codec algorithms for encoding a self-inverting permutation $\pi^*$ into
two different reducible permutation flow-graphs $F_1[\pi^*]$ and $F_2[\pi^*]$ having properties deriving from the bitonic subsequences composing
the self-inverting permutation $\pi^*$. We show that the permutation $\pi^*$ can be efficiently encoded into either $F_1[\pi^*]$ or $F_2[\pi^*]$ and also correctly and efficiently extracted from theses flow-graphs. Finally, in Section~5 we conclude the paper and discuss possible future extensions.

\section{Theoretical Framework}
\vspace*{0.05in}

We consider finite graphs with no multiple edges. For a graph~$G$, we denote by $V(G)$ and $E(G)$ the vertex (or, node) set and edge set of $G$, respectively. The subgraph of a graph $G$ induced by a set $S \subseteq V(G)$ is denoted by $G[S]$. The \emph{neighborhood}~$N(x)$ of a vertex~$u$ of the graph~$G$ is the set of all the vertices of $G$ which are adjacent to $u$.
The \emph{degree} of a vertex~$u$ in the graph~$G$, denoted $deg(u)$, is the number of edges incident on node $u$; for a node $u$ of a directed graph $G$, the number of head-endpoints of the directed edges adjacent to $u$ is called the indegree of the node $u$, denoted $indeg(u)$, and the number of tail-endpoints is its outdegree, denoted $outdeg(u)$. The parent of a node~$x$ of a rooted tree $T$ is denoted by $p(x)$.

\subsection{Previous Results}
\vspace*{0.05in}
In mathematics, the notion of permutation relates to the act of arranging all the members of a set into a sequence or order. Permutations may be represented in many ways \cite{Book-SF96}, where the most straightforward is simply a rearrangement of the elements of the set $N_n = \{1, 2, \ldots, n\}$. For example, $\pi = (5, 6, 8, 9, 1, 2, 7, 3, 4)$ is a permutation of the elements of the set $N_{9}$; hereafter, we shall say that $\pi$ is a permutation over the set $N_{9}$.

\vspace*{0.15in}
\noindent {\bf Definition~2.1} Let $\pi=(\pi_1, \pi_2, \ldots, \pi_n)$ be a permutation over the set $N_{n}$, $n>1$. The inverse of the permutation $\pi$ is the permutation $q=(q_1, q_2, \ldots, q_n)$ with $q_{\pi_i} = \pi_{q_i} = i$. A {\it self-inverting permutation} (or, for short, SiP) is a permutation that is its own inverse: $\pi_{\pi_i} = i$.

\vspace*{0.15in}
\noindent By definition, a permutation is a SiP (self-inverting permutation) if and only if all its cycles are of length 1 or 2; for example, the permutation $\pi = (5, 6, 8, 9, 1, 2, 7, 3, 4)$ is a SiP with cycles: $(1,5)$, $(2,6)$, $(3,8)$, $(4,9)$, and $(7)$.
Throughout the paper we shall denote a self-inverting permutation $\pi$ over the set $N_n$ as $\pi^*$.

A flow-graph is a directed graph $F$ with an initial node $s$ from which all other nodes are reachable. A directed graph $G$ is strongly connected when there is a path $x \rightarrow y$ for all nodes $x$, $y$ in $V(G)$. A node $u \in V(G)$ is an {\it entry} for a subgraph $H$ of the graph $G$ when there is a path $p = (y_1, y_2, \ldots, y_k, x)$ such that $p \cap H = \{x\}$ (see, \cite{HU72,HU74}).

\vspace*{0.15in}
\noindent {\bf Definition~2.2} A flow-graph is reducible when it does not have a strongly connected subgraph with two (or more) entries.

\vspace*{0.15in}
\noindent There are some other equivalent definitions of the reducible flow-graphs which use a few more graph-theoretic concepts. A depth first search (DFS) of a flow-graph partitions its edges into tree, forward, back, and cross edges. It is well known that tree, forward, and cross edges form a dag known as a DFS dag. Hecht and Ullman show that a flow-graph $F$ is reducible if and only if $F$ has a unique DFS dag \cite{HU72,HU74}.

Recently, a wide range of software watermarking techniques has been proposed among which the graph-based methods that encode watermark numbers $w$ as reducible flow-graph structures $F$ capturing such properties which make them resilient to attacks.

In \cite{CN10}, Chroni and Nikolopoulos presented the encoding algorithm ${\tt Encode\_W.to.SIP}$, along with its corresponding decoding one, which encodes a watermark number $w$ as a self-inverting permutation $\pi^*$ in $O(n)$ time, where $n$ is the length of the binary representation of the integer $w$. Later, the same authors introduced a type of reducible permutation graphs $F[\pi^*]$ and  proposed several efficient codec methods which embed a self-inverting permutation $\pi^*$ into such a reducible permutation graph $F[\pi^*]$ and efficiently extract it from the graph $F[\pi^*]$. Moreover, they proposed several algorithms for multiple encoding the same watermark number $w$ into many different reducible permutation graphs $F_i[\pi^*]$, $i > 1$, through the use of the encoding self-inverting permutation $\pi^*$ \cite{CN11,CN15}.


These results are summarized in the following theorems and lemmata.

\vspace*{0.15in}
\noindent {\bf Theorem~2.1} {\it Let $w$ be an integer and let $b_1b_2\cdots b_n$ be
the binary representation of $w$. The number~$w$ can be encoded into a self-inverting permutation~$\pi^*$ of length $2n+1$
and correctly extracted from $\pi^*$ in $O(n)$ time and space.}

\vspace*{0.15in}
\noindent {\bf Theorem~2.2} {\it Let $\pi^*$ be a self-inverting permutation of length $n^*$ which encodes a watermark integer $w$. The permutation $\pi^*$ can be encoded as
a reducible permutation flow-graph $F[\pi^*]$ and correctly extracted from $F[\pi^*]$ in $O(n^*)$ time and space.}

\vspace*{0.15in}
\noindent {\bf Lemma~2.1} {\it Let $F[\pi^*]$ be a reducible permutation graph of size $O(n^*)$
constructed by a proper encoding algorithm. The unique Hamiltonian path of $F[\pi^*]$ can be computed in $O(n^*)$ time and space.}

\vspace*{0.15in}
\noindent {\bf Theorem~2.3} {\it We can produce more than one reducible flow-graphs $F_1[\pi^*], F_2[\pi^*], \ldots, F_n[\pi^*]$ which encode the
same watermark integer $w$ through the use of the self-inverting permutation $\pi^*$.}

\vspace*{0.1in}
\subsection{Bitonic Sequences and SiPs}
\label{subsec:Bitonic-Sequences}
\vspace*{0.05in}
A sequence $b=(b_1, b_2, \ldots, b_n)$ is called bitonic if either monotonically increases and then monotonically decreases, or else monotonically decreases and then monotonically increases.

In this paper, we consider only bitonic sequences that monotonically increases and then monotonically decreases, i.e., the minimum element of such a sequence $b$ is either the first $b_1$ or the last $b_n$ element of $b$; for example, $b=(5, 6, 8, 9, 1)$ is such a bitonic sequence. The maximum element of a bitonic sequence $b$, which we call {\it top} element of $b$, is denoted as $top(b)$. Obviously, $b$ is an increasing sequence if $top(b)=b_n$, while $b$ is a decreasing sequence if $top(b)=b_1$.

\vspace*{0.15in}
\noindent {\bf Definition~2.3} Let $b=(b_1, b_2, \ldots, b_n)$ be a bitonic sequence of length $n$. According to the index of the element $top(b)$, the sequence $b$ is called:

\begin{itemize}
\vspace*{0.0in}
\item[$\circ$] $i$-bitonic or {\it increasing bitonic} if $top(b)=b_n$;
\vspace*{0.0in}
\item[$\circ$] $d$-bitonic or {\it decreasing bitonic} if $top(b)=b_1$;
\vspace*{0.0in}
\item[$\circ$] $id$-bitonic or {\it full-bitonic} if $b_1 < top(b)$ and $top(b) > b_n$.
\end{itemize}

\vspace*{0.1in}
\noindent In terms of the above definition, $b_1=(2,7)$ is an $i$-bitonic or increasing bitonic sequence, $b_2=(4,3)$ is a $d$-bitonic or decreasing bitonic sequence, while $b_3=(5, 6, 8, 9, 1)$ is an $id$-bitonic or full-bitonic sequence.

Let $\pi^*$ be the self-inverting permutation of length $n^*$ produced by the encoding algorithm ${\tt Encode\_W.to.SIP}$ \cite{CN10}, and let $b_1^*, b_2^*, \ldots, b_k^*$ be the bitonic subsequences forming the permutation $\pi^*$; note that, $\pi^*$ encodes a watermark number $w$ of binary length $n$ and $n^*=2n+1$. Then, $b_1^*, b_2^*, \ldots, b_k^*$ have the following properties:
\begin{itemize}
\vspace*{0.0in}
\item[$(P_1)$] Sequence $b_1^*$ is a full-bitonic, $b_2^*, b_3^*, \ldots, b_{k-1}^*$ are either full-bitonic or $d$-bitonic sequences, while $b_k^*$ is either a full-bitonic, $i$-bitonic, or $d$-bitonic sequence.
\vspace*{0.0in}
\item[$(P_2)$] Sequence $b_1^*$ contains the max element $\pi^*_{max}$ and the min element $\pi^*_{min}$ of permutation $\pi^*$ and has always length $\lceil n^*/2 \rceil$; note that, $\pi^*_{max}=top(b^*_1)=2n+1$ and $\pi^*_{min}=1$;
\vspace*{0.0in}
\item[$(P_3)$] The last element $last(b^*_k)$ of sequence $b_k^*$ is equal to the index of the max element $\pi^*_{max}=2n+1$ in $b_1^*$.
\end{itemize}

\noindent {\bf Example 2.1} Let $w_1=20$ and $w_2=45$ be two watermark numbers. For these two watermarks, the encoding algorithm ${\tt Encode\_W.to.SIP}$ produces the self-inverting permutations
\begin{itemize}
\vspace*{0.0in}
\item[$\circ$\,] $\pi^*_1=(6,8,11,10,9,1,7,2,5,4,3)$, and
\vspace*{0.0in}
\item[$\circ$\,] $\pi^*_2=(7,9,10,12,13,11,1,8,2,3,6,4,5)$
\end{itemize}

\vspace*{0.0in}
\noindent of lengths $n^*_1=2n_1+1=11$ and $n^*_2=2n_2+1=13$, respectively; note that, $n_1=5$ is the length of the binary representation of number 20 (i.e., 10100), while $n_2=6$ is that of number 44 (i.e., 101101).
The permutations $\pi^*_1$ and $\pi^*_2$ are composed by the following three and four, respectively, bitonic subsequences:
\begin{itemize}
\vspace*{0.0in}
\item[$\circ$\,] $\pi^*_1$ $: \ $ $(6,8,11,10,9,1) \ || \ (7,2) \ || \ (5,4,3)$
\vspace*{0.0in}
\item[$\circ$\,] $\pi^*_2$ $: \ $ $(7,9,10,12,13,11,1) \ || \ (8,2) \ || \ (3,6,4) \ || \ (5)$
\end{itemize}

\vspace*{0.15in}
\noindent We observe that all the bitonic subsequences of both permutations $\pi^*_1$ and $\pi^*_2$ satisfy the properties $(P_1)$, $(P_2)$, and $(P_3)$. Indeed, for example, the subsequence $b_1^*=(6,8,11,10,9,1)$ of permutation $\pi^*_1$ is full-bitonic, contains the max and the min elements of $\pi^*_1$, has length $6=\lceil n^*_1/2 \rceil$, where $n^*_1=11$ is the length of $\pi^*_1$, and the last element of subsequence $b_3^*=(5,4,3)$ (i.e., 3) is equal to the index (i.e., 3) of the max element of $\pi^*_1$ in sequence $b_1^*$ (i.e., $\pi^*_{max}=11$).

\vspace*{0.1in}
\noindent {\bf Example 2.2} Here is an example of the permutation $\pi^*$ which encodes the number $w=54$ with binary representation $110110$, i.e., $\pi^*=(7,8,10,11,13,12,1,2,9,3,4,6,5)$. It is easy to see that this permutation satisfies the properties $(P_1)-(P_3)$ and all its bitonic subsequences are of type full-bitonic; indeed, $b_3^*=(7,8,10,11,13,12,1)$, $b_3^*=(2,9,3)$, and $b_3^*=(4,6,5)$.

\vspace*{0.2in}
\section{Bitonic Algorithm}
\vspace*{0.08in}
Having presented an efficient codec algorithm for encoding a watermark number $w$ as a self-inverting permutation $\pi^*$ \cite{CN10} and several codec algorithms for efficiently encoding the permutation $\pi^*$ into different reducible permutation flow-graphs $F_i[\pi^*]$ $(i > 1)$, in this section we extend the types of such flow-graphs by proposing an algorithm for encoding a self-inverting permutation $\pi^*$ into a reducible permutation graph $F[\pi^*]$ having properties which are derived from the bitonic subsequences composing
the self-inverting permutation $\pi^*$ (see, properties $P_1-P_3$).

The encoding algorithm, which we call {\tt Encode\_SIP.to.RPG-Bitonic-1} is described below.

\vspace*{0.2in}
\noindent {\bf Algorithm {\tt Encode\_SIP.to.RPG-Bitonic-1}}
\vspace*{0.0in}
\begin{itemize}
\item[$1.$\,] Compute the bitonic subsequences $S_1, S_2, ..., S_k$ of the self-inverting permutation $\pi^*$ and
let $S_i=(i_1, i_2, ..., top(S_i), ...,i_t)$;
\item[$2.$\,] Construct a directed graph $F_1[\pi^*]$ on $n^*+2$ vertices as follows:
\begin{itemize}
\item[$2.1$\,] $V(F_1[\pi^*])=\{s=u_{n^*+1},  u_{n^*}, ..., u_1, u_0=t\}$;
\item[$2.2$\,] for $i=n^*, n^*-1, ...,0$ do

\hspace{0.5cm} add the edge $(u_{i+1}, u_i)$ in $E(F_1[\pi^*]$);

\end{itemize}
\item[$3.$\,] For each bitonic subsequence $S_i$, $1\le i\le k$, do
\begin{itemize}

\item[$3.1$\,] add the edge $(u_{top({S_i})}, s)$ in $E(F_1[\pi^*])$;

\item[$3.2$\,] for $j=1, 2, \ldots, top(S_i), \ldots, t-1$ do

\begin{itemize}
\item[] if $i_j < i_{j+1}$ then add the edge $(u_{i_j}, u_{i_{j+1}})$ \\ else the edge $(u_{i_{j+1}}, u_{i_j})$ in $E(F_1[\pi^*])$;
\end{itemize}

\end{itemize}

\item[$4.$\,] Return the graph $F_1[\pi^*]$;
\end{itemize}

\begin{figure}[t!]
    \hrule\medskip\medskip\smallskip
    \centering
    \includegraphics[scale=0.55]{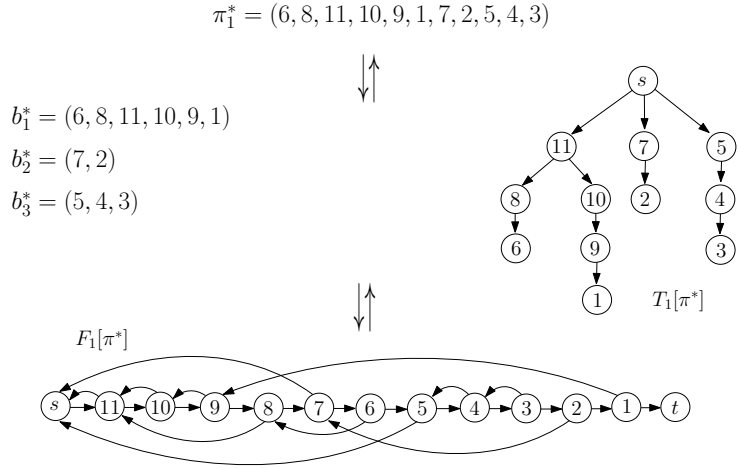}
    \centering
    \smallskip\medskip\hrule\medskip
    \caption{\small{The main structures used or constructed by the codec algorithms {\tt Encode\_SiP.to.RPG-Bitonic-1} and
{\tt Decode\_RPG.to.SiP-Bitonic-1}.}}
\label{fig:fig-1-20}
\end{figure}

\vspace*{0.1in}
\noindent Figure~\ref{fig:fig-1-20} shows the encoding of the self-inverting permutation $\pi^*=(6,8,11,10,9,1,7,2,5,4,3)$ into the reducible permutation flow-graph $F_1[\pi^*]$; note that, $\pi^*$ encodes the watermark number $w=20$.

Let $w$ be a watermark number and $\pi^*$ be the self-inverting permutation of length $n^*=2n+1$ which encodes watermark $w$, where $n$ is the length of the binary representation of number $w$. Analyzing the time and space performance of our encoding algorithm {\tt Encode\_SIP.to.RPG-Bitonic-1}, we can obtain the following result:

\vspace*{0.15in}
\noindent {\bf Theorem~3.1} {\it The algorithm {\tt Encode\_SIP.to.RPG-Bitonic-1} for encoding the permutation $\pi^*$ as a reducible permutation flow-graph $F_1[\pi^*]$ requires $O(n)$ time and space.}

\vspace*{0.15in}
\noindent We next present the decoding algorithm {\tt Decode\_RPG.to.SIP-Bitonic-1}, which takes as input a reducible permutation flow-graph $F_1[\pi^*]$ on $n^*+2$ nodes produced by algorithm {\tt Encode\_SIP.to.RPG-Bitonic-1} and correctly extract the self-inverting permutation $\pi^*$ from the graph $F_1[\pi^*]$; the algorithm is described below.

\vspace*{0.2in}
\noindent {\bf Algorithm {\tt Decode\_RPG.to.SIP-Bitonic-1}}
\vspace*{0.0in}
\begin{itemize}
\item[1.\,]
Delete the directed edges $(u_{i+1}, u_i)$, $1 \leq i \leq n$,
and the node $t = u_0$ from $F_1[\pi^*]$, and flip all the remaining directed edges in $F_1[\pi^*]$; let $s=u_0, u_1, u_2, \ldots, u_n$ be the nodes in the resulting graph $T_1[\pi^*]$;
\vspace*{0.0in}
\item[$2.$\,] Compute the set $R=\{u_j \ | \ (s,u_j) \in T_1[\pi^*]\}$ and delete the directed edges $(s,u_j)$ from the graph $T_1[\pi^*]$;

\item[$3.$\,] Sort the nodes of set $R$ in descending order according to their labels and let $R^*=(r_1,r_2,...,r_k)$ be the resulting sorted sequence, $1 \le k < n$;

\item[$4.$\,] Construct the underlying graph $H[\pi^*]$ of the directed graph $T_1[\pi^*]$ and let $C(r_1)$, $C(r_2)$, $\ldots$, $C(r_k)$ be the connected components of the graph $H[\pi^*]$ which contain the nodes $r_1,r_2,...,r_k$, respectively;

\item[$5.$\,] For each node $r_i \in R^*$, $i=1,2,...,k$, perform BFS-search in graph $C(r_i)$ starting at node $u$ and compute the sequence $b_i^*$ of the nodes of $C(r_i)$ taken by the order in which they are BFS-discovered; the starting node $u$ is selected as follows:
\begin{itemize}
\item[$\circ$\,] if $i<k$ and $deg(r_i)=2$, then $u$ is the node with minimum label in $C(r_i)$;
\item[$\circ$\,] if $i<k$ and $deg(r_i)\le1$, then $u$ is the node with maximum label in $C(r_i)$;
\item[$\circ$\,] if $i=k$, then $u$ is the node with label $\ell_{max}$ in $C(r_i)$, where $\ell_{max}$ is the
index of the max element in sequence $(b_1^*)^R$;
\end{itemize}

recall that, $(b_i^*)^R$ denotes the reverse sequence of $b_i^*$, $1 \leq i \leq k$;

\item[$6.$\,] Return $\pi^*=(b_1^*)^R || b_2^* || \cdots || b_{k-1}^* || (b_k^*)^R$;

\end{itemize}

\vspace*{0.1in}
\noindent Figure~\ref{fig:fig-1-20} shows the extraction of the self-inverting permutation $\pi^*=(6,8,11,10,9,1,7,2,5,4,3)$, which encodes the watermark number $w=20$, from the reducible permutation flow-graph $F_1[\pi^*]$ using the tree $T_1[\pi^*]$. Note that, a reducible permutation flow-graph $F_1[\pi^*]$ of size $n^*+2$ encodes a self-inverting permutation $\pi^*$ of length $n^*$.

The following result summarizes the correctness and the time and space complexity of our proposed decoding algorithm {\tt Decode\_RPG.to.SIP-Bitonic-1}.

\vspace*{0.15in}
\noindent {\bf Theorem~3.2} {\it The algorithm {\tt Decode\_RPG.to.SIP-Bitonic-1} correctly extract the permutation $\pi^*$ from a reducible permutation flow-graph $F_1[\pi^*]$ in $O(n)$ time and space.}

\vspace*{0.1in}
\section{Full-Bitonic Algorithm}
\vspace*{0.05in}
In this section we enrich the repository of reducible permutation flow-graphs $F[\pi^*]$ which can encode a self-inverting permutation $\pi^*$ or, equivalently, a watermark number $w$ by proposing a reducible permutation flow-graph $F_2[\pi^*]$, different from $F_1[\pi^*]$ but of the same type, having also important properties deriving from the bitonic subsequences of $\pi^*$.

By construction, the indegree of the first node $s=u_{n^*+1}$ of the flow-graph $F_1[\pi^*]$ is equal to the number of bitonic subsequences $b_1^*, b_2^*, \cdots, b_k^*$ of $\pi^*$, while the indegree of the first node of the graph $F_2[\pi^*]$ is much smaller that $k$. This property causes $F_2[\pi^*]$ more appropriate, in same cases, since it does not contain an extreme characteristic thereby enhancing the resilience of graph-structure to attacks.
The proposed algorithm {\tt Encode\_SIP.to.RPG-Bitonic-2} for encoding a self-inverting permutation $\pi^*$ into a reducible permutation graph $F_2[\pi^*]$ is described below.

\begin{figure*}[t!]
    \hrule\medskip\medskip\smallskip
    \centering
    \includegraphics[scale=0.52]{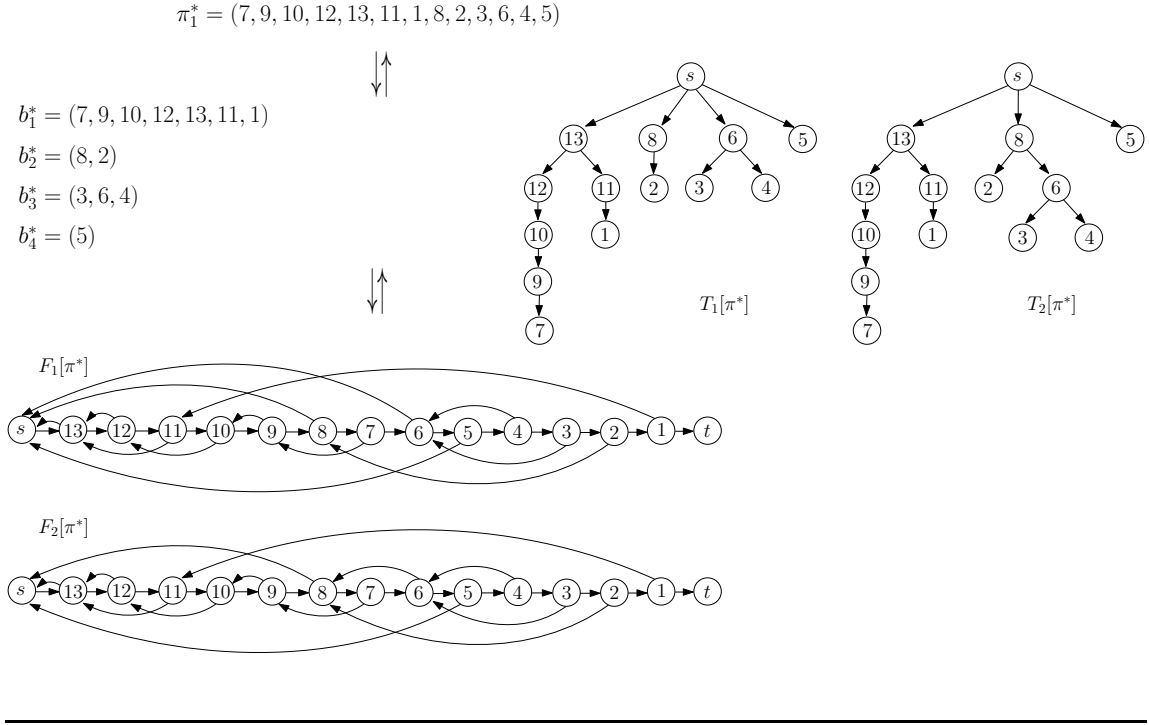}
    \centering
    \medskip\medskip\hrule\medskip
    \caption{\small{The main structures used or constructed by the codec algorithms {\tt Encode\_SiP.to.RPG-Bitonic-2} and
{\tt Decode\_RPG.to.SiP-Bitonic-2}.}}
\label{fig:fig-1-45}
\end{figure*}

\vspace*{0.2in}
\noindent {\bf Algorithm {\tt Encode\_SIP.to.RPG-Bitonic-2}}
\vspace*{0.0in}
\begin{itemize}
\item[$1.$\,] Execute algorithm {\tt Encode\_SIP.to.RPG-Bitonic-1} and compute the bitonic subsequences
$S_1, S_2, ..., S_k$ of $\pi^*$ and the graph $F_1[\pi^*]$; Set $F_2[\pi^*] \leftarrow F_1[\pi^*]$;
\item[$2.$\,] For each edge $(u_{top({S_i})}, s)$ in $F_2[\pi^*]$, $2 \leq i \leq k$, do

if $S_i$ is a full-bitonic sequence, then

\begin{itemize}
\item[$\circ$\,] delete the edge $(u_{top({S_i})}, s)$ and

\item[$\circ$\,] add the edge $(u_{top({S_i})}, u_{top({S_{i-1}})})$ in $E(F_2[\pi^*])$;
\end{itemize}

\item[$3.$\,] Return the graph $F_2[\pi^*]$;
\end{itemize}

\vspace*{0.1in}
\noindent We next describe the corresponding decoding algorithm for extracting the permutation $\pi^*$ from the flow-graph $F_2[\pi^*]$.

\vspace*{0.2in}
\noindent {\bf Algorithm {\tt Decode\_RPG.to.SIP-Bitonic-2}}
\vspace*{0.0in}
\begin{itemize}

\item[$1.$\,] Execute Steps 1 and 2 of algorithm {\tt Decode\_RPG.to.SIP-Bitonic-1} on graph $F_2[\pi^*]$ and compute the directed graph $T_2[\pi^*]$ and the node set $R$;

\item[$2.$\,] Compute the node set

$R'=\{u_j \ | \ (u_i,u_j) \in T_2[\pi^*] \ \text{and} \ outdeg(u_j) \geq 2\}$,

delete the directed edges $(u_i,u_j)$ from the graph $T_2[\pi^*]$, and set $R \leftarrow R \cup R'$;

\item[3.\,] Execute Steps 3, 4 and 5 of the decode algorithm {\tt Decode\_RPG.to.SIP-Bitonic-1} and compute the sequences $b_1^*, b_2^*, \ldots, b_k^*$;

\item[$4.$\,] Return $\pi^*=(b_1^*)^R || b_2^* || \cdots || b_{k-1}^* || (b_k^*)^R$;

\end{itemize}

\vspace*{0.1in}
\noindent In the example of Figure~2, $R=\{13,8,5\}$ and $R'=\{6\}$. Recall that, the self-inverting permutation which encodes watermark $w$ is of length $n^*=2n+1$, where $n$ is the binary length of the watermark number $w$, while the reducible permutation flow-graph $F_2[\pi^*]$ is of size $n^*+2$.

The results of this section concerning the correctness and the time and space complexity of both algorithms are summarized in the following theorem.

\vspace*{0.15in}
\noindent {\bf Theorem~4.1} {\it The algorithm {\tt Encode\_SIP.to.RPG-Bitonic-2} encodes a permutation $\pi^*$ into a reducible permutation flow-graph $F_2[\pi^*]$ in $O(n)$ time and space and the corresponding decoding algorithm {\tt Decode\_RPG.to.SIP-Bitonic-2} correctly extract $\pi^*$ from the flow-graph $F_2[\pi^*]$ within the same time and space complexity.}

\vspace*{0.1in}
\section{Concluding Remarks}
\vspace*{0.05in}

In the last decade, a wide range of software watermarking techniques has been proposed among which the graph-based methods that encode watermark numbers as graphs whose structure resembles that of real program graphs. Recently, Chroni and Nikolopoulos \cite{CN11,CN15} proposed several algorithms for multiple encoding a watermark into a graph-structure: an integer (i.e., a watermark) is encoded first into a self-inverting permutation $\pi^*$ and then into several reducible permutation graphs using Cographs \cite{CN11} and Heap-ordered trees \cite{CN15}.

Following up on our recently proposed methods, in this paper, we extended the class of graph-structures by proposing two different reducible permutation flow-graphs $F_1[\pi^*]$ and $F_2[\pi^*]$ incorporating important structural properties which are derived from the bitonic subsequences forming the self-inverting permutation $\pi^*$. These new flow-graphs enrich the repository of graphs available for multiple encoding a watermark number and, thus, it increases our ability to select a graph structure more similar to the structure of a given application program $P$ thereby enhancing the resilience of our codec system to attacks.

An interesting open question is whether the properties of the bitonic subsequences forming the self-inverting permutation $\pi^*$ can help develop efficient graph structures having more similar structure to that of a given application program $P$. Can we use some of the bitonic subsequences $b_1^*, b_2^*, \cdots, b_k^*$ of permutation $\pi^*$ or part of them in order to efficiently encode and decode a self-inverting permutation $\pi^*$ into a reducible permutation flow-graph $F[\pi^*]$? we leave it as an open problem for future investigation.

\end{document}